\documentstyle[fleqn,espcrc2,epsf]{article}

\newcommand{\AmS}{{\protect\the\textfont2
  A\kern-.1667em\lower.5ex\hbox{M}\kern-.125emS}}
\title{One-flavour QCD at finite temperature
\thanks{Talk  presented by C. Alexandrou }}

\author{Constantia Alexandrou\address{Department of Natural Sciences, 
        University of Cyprus, CY-1678 Nicosia, Cyprus},
        Artan Bori\c{c}i\address{Paul Scherrer Institute, 
CH-5232 Villigen, Switzerland},
        Alessandra Feo$^{\rm a}$,
        Philippe de Forcrand\address{Swiss Center for Scientific Computing, 
        ETH-Zentrum, CH-8092 Z\"urich, Switzerland},
        Andrea Galli\address{ELCA Informatique, HofwiesenStr. 26, 
        CH-8057 Z\"urich, Switzerland},
        Fred Jegerlehner\address{DESY-IfH Zeuthen, D-15738 Zeuthen, Germany},
        and 
        Tetsuya Takaishi\address{Hiroshima University of Economics,
Hiroshima, Japan 731-01}
}

\begin{document}

\begin{abstract}
We present results, for heavy to moderate quark masses, of a study
of thermodynamic properties of 1-flavour QCD, using the multiboson algorithm.
Finite-size scaling behaviour is studied on lattices of size 
$8^3\times 4$, $12^3\times 4$ and $16^3\times 4$. It is shown that,
for heavy quarks,
the peak of the Polyakov loop susceptibility grows linearly
with the spatial volume, indicating a first order phase
transition. The deconfinement ratio and the distribution of the norm of
the Polyakov loop corroborate this result.
For moderately heavy quarks the first-order transition weakens and becomes
a crossover.
We estimate the end point of the first-order phase
transition to occur at a quark mass of about 1.6 GeV. 
\end{abstract}

\maketitle

\section{Introduction}
 
We use the non-hermitian variant
of the multiboson method \cite{Luescher}
to simulate one-flavour QCD at finite temperature 
(for more details on the algorithm see \cite{lat96,BPhG}).
We perform a finite-size scaling study for four $\kappa$ values of Wilson 
fermions.
Because we simulate heavy quarks only, we can safely ignore the
problem of the sign of the fermionic determinant.
Global observables, such as the Polyakov loop, obtained with our
multiboson algorithm showed an autocorrelation time ${\cal O}(10)$ shorter
than if obtained with an efficient polynomial variant of HMC,
appropriate to simulate 1 flavour \cite{lat96_TT}. 
This substantial improvement can presumably be attributed 
to the additional freedom due to
the bosonic fields of the multiboson algorithm enabling movement 
{\em around} energy
barriers in constrast to HMC where one has to go {\em over} the barrier.

In Table~1 we collect the parameters of our simulations.

\begin{table}[t]
\caption{Summary of simulations:
for each $\kappa$ value and volume studied,
$N_b$ indicates the number of bosonic
fields used and $acc$ the average acceptance; $Ksw$ is the total number 
of thermalized configurations (in thousands) used in the
reweighting procedure. ~~~~~~~~~~~\label{table_parameters}}

\small
\begin{center}
\begin{tabular}{|c|@{~}c@{~}c|@{~}c@{~}c|@{~}c@{~}c|}
%\begin{tabular}{|c|cc|cc|cc|}
\hline
$\kappa$ &\multicolumn{2}{c}{ $8^3 \times 4$} &
    \multicolumn{2}{c}{$ 12^3 \times 4$} & 
\multicolumn{2}{c|}{$ 16^3 \times 4 $ }\\
\hline
 & $N_b/acc$ & $Ksw$ & $N_b/acc$ & $Ksw$ & $N_b/acc$ & $Ksw$ \\
\hline
 0.05 & 8/0.78 & 18 & 12/0.74 &20 & 24/0.83 & 20 \\
\hline
0.10 & 16/0.67 & 45 & 24/0.63 & 50 & 32/0.67 & 27*\\
\hline
0.12 & 24/0.74 &55 & 32/0.67 & 40& 40/0.69 & 10*\\
\hline
0.14 & 32/0.77 & 60 & 40/0.70 &37 & 50/0.67 & 12\\
\hline
\end{tabular}
\end{center}
{~ \\ {$~^\star$} {\footnotesize 
indicates that we are still accumulating statistics.}}
\end{table}
\normalsize

\section{Determination of $\beta_c$}

To identify the critical line $\beta(\kappa)$ we study the
signals for a first order phase transition on a finite lattice
(e.g. \cite{1rst}) using three observables:
\begin{itemize}
\item 
 The histogram of the norm of the Polyakov loop, which shows 2 peaks
of equal area at
criticality. For the lattices under study here 
we observed enough  tunneling events between
the two phases to make this a reliable method. This is shown in Fig.~1
for $\kappa = 0.05$ for the three lattice sizes.

\item The deconfinement ratio $\rho \equiv 3/2 p - 1/2$, where $p$ is
the
probability for the complex trace of the Polyakov loop to fall within
$\pm 20 \deg$ of a $Z_3$ axis. $\rho$ goes from 0 (uniform angular distribution)
to 1 (alignment with $Z_3$ axis).

\item The peak value of the  
 susceptibility of the Polyakov loop.
\end{itemize}

Using these criteria we determined the critical values $\beta(\kappa)$ given in
Table~2.

\begin{table}[h]
\vspace{-0.8cm}
\begin{center}
\begin{tabular}{|c|c|c|c|c|}
\hline
$\kappa$ & 0.05 &0.10 &0.12&0.14  \\ \hline
$\beta_c$ &5.692 & 5.66   & 5.63 & 5.59 \\ \hline
\end{tabular}
\end{center}
\caption{
\label{table_results}
The critical values $\beta(\kappa)$. The estimated error is about one 
on the last digit. }
\vspace{1ex}
\end{table}
\vspace*{-1.0cm}

\begin{figure}[t]
\begin{center}
\vspace*{-1.0cm}
 \mbox{\epsfxsize=8cm\epsfysize=12cm\epsffile{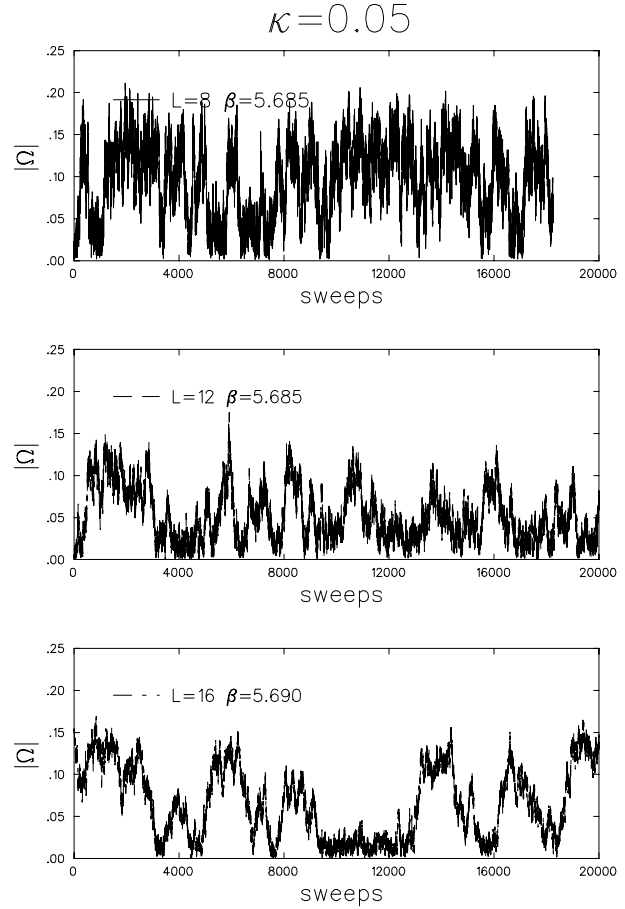}}
\end{center}
\vspace{-1.2cm}
\caption{Tunneling between the two phases for $\protect\kappa=0.05$.
By $|\Omega|$ we denote the norm of the spatial average of the
Polyakov line.} 
\end{figure}

\section{Order of transition}
In Fig.~2 we show the deconfinement ratio $\rho$ obtained using
reweighting~\cite{rew}
for $\kappa=0.05, \kappa=0.10$ and 
$\kappa=0.12$. Across the transition region the slope of $\rho(\beta)$
increases with the volume for $\kappa=0.05$ signaling a first order
transition.
This is to be contrasted with the behaviour of $\rho$ for
$\kappa=0.12$ indicating that $\kappa=0.12$ is
already in the crossover region. For $\kappa=0.10$ we observe a 
 qualitative behaviour more appropriate for a weak first order transition.

In Fig.~3 we display the peak of the susceptibility versus the spatial volume $V$. The
lines shown are best fits to the form $V^{\alpha}$. For $\kappa=0.05$ 
the best fit yields $\alpha=0.96(4)$ whereas for $\kappa=0.14$ 
$\alpha$ is zero. For $\kappa=0.10$ and $\kappa=0.12$ the situation is
less clear. The small value of $\alpha=0.24(4)$ at $\kappa=0.12$
as well as the absence of tunneling indicate that we are in the
crossover region. 
For $\kappa=0.10$ $\alpha=0.54(3)$,
increasing with the statistics of the $L=16$ lattice (still in progress),
and tunneling is still observed. If this increasing trend of the peak of
the susceptibility for $L=16$  continues then we will have confirmed
that the transition is still first order. 

Considering the signals that we have for $\kappa=0.1$, namely
tunneling between the two phases, the behaviour of the deconfinement
ratio and the finite size scaling of the susceptibility, we
are  led to the conclusion that the end point of the first
order phase transition occurs very near $\kappa=0.1$.

Taking the end-point value of $\kappa$, $\kappa_{ep} \approx 0.1 \pm 0.01$,
we can approximately
map to physical units, using the tadpole-improvement property 
$\kappa_c(\beta) \langle\Box\rangle^{1/4} \approx 1/8$ to obtain 
$m_q a \sim 1.8(5)$, and $(4 a)^{-1} \sim 220 MeV$ from the deconfinement 
temperature. This gives $m_q \sim 1.6(5) GeV$ at the end-point.

\begin{figure}[t]
\begin{center}
\vspace{-0.7cm}
 \mbox{\epsfxsize=8cm\epsfysize=13cm\epsffile{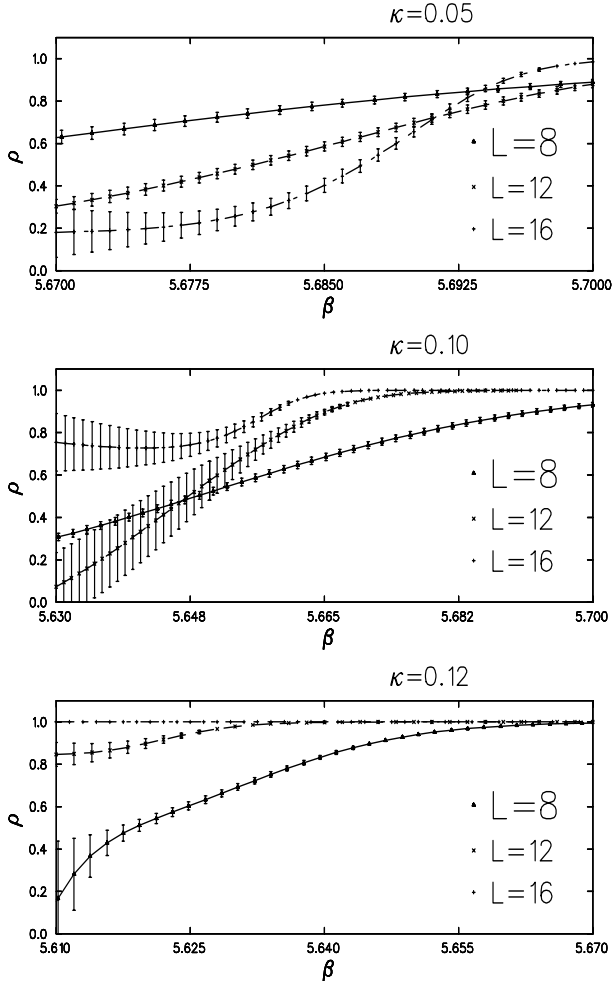}}
\end{center}
\vspace*{-1.5cm}
\caption{Deconfinement ratio for $\kappa=0.05$, $\kappa=0.10$ and $\kappa=0.12$ for
three lattice sizes}
\end{figure}

\begin{figure}[t]
\begin{center}
\vspace{-0.5cm}
 \mbox{\epsfxsize=8cm\epsfysize=8cm\epsffile{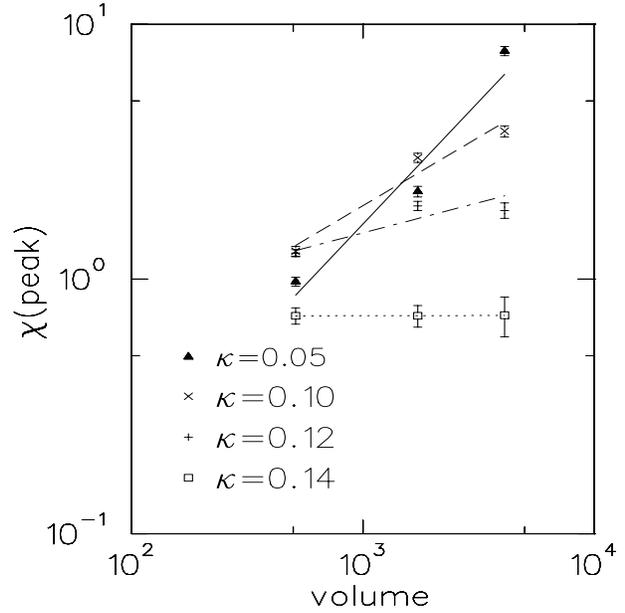}}
\end{center}
\vspace*{-1.0cm}
\caption{Volume dependence of the maximum of the Polyakov loop susceptibility}
\end{figure}

This is in line with phenomenological expectations. The pure gauge deconfinement
transition is fairly weak, with a correlation length 
$\cal O$(a few $\sigma^{-1/2}$). This is the minimum system size necessary
to observe the deconfinement transition.
Dynamical quarks introduce a new length
scale, the distance where the string breaks, 
${\cal O}(2 m_q /\sigma)$. Confinement can only be observed up to
this distance. 
When the quark mass is lowered sufficiently that the second length-scale 
is similar to (or smaller than) the first, one cannot tell if the system
is confined or deconfined, and the transition is replaced by a crossover.
This occurs for
$m_q \sim {\cal O}(a few \sqrt\sigma / 2)$, i.e. $m_q \sim {\cal O}(1) GeV$.

\vspace{0.2cm}
{\bf Acknowledgements:} We thank SIC of the EPFL in
Lausanne, ZIB in Berlin and  the Supercomputing Institute of the
University of Minnesota for computer time.

%\vspace{-0.3cm}

\end{document}